\documentclass[conference]{IEEEtran}
\IEEEoverridecommandlockouts
\usepackage{cite}
\usepackage{amsmath,amssymb,amsfonts}
\usepackage{algorithmic}
\usepackage{graphicx}
\usepackage{textcomp}
\usepackage{xcolor}
\usepackage{hyperref}
\def\BibTeX{{\rm B\kern-.05em{\sc i\kern-.025em b}\kern-.08em
    T\kern-.1667em\lower.7ex\hbox{E}\kern-.125emX}}
\begin{document}

\title{Persona-Based Requirements Engineering for Explainable Multi-Agent Educational Systems: A Scenario Simulator for Clinical Reasoning Training
}

\author{\IEEEauthorblockN{1\textsuperscript{st} Weibing Zheng}
\IEEEauthorblockA{\textit{School of Information Technology} \\
\textit{University of Cincinnati, OH }\\
Cincinnati, USA \\
zhengwb@mail.uc.edu}
\and
\IEEEauthorblockN{2\textsuperscript{nd} Laurah Turner}
\IEEEauthorblockA{\textit{College of Medicine} \\
\textit{University of Cincinnati, OH }\\
Cincinnati, USA \\
turnela@ucmail.uc.edu}

\and
\IEEEauthorblockN{3\textsuperscript{rd} Jess Kropczynski}
\IEEEauthorblockA{\textit{School of Information Technology} \\
\textit{University of Cincinnati, OH }\\
Cincinnati, USA \\
kropczjn@ucmail.uc.edu}

\and
\IEEEauthorblockN{4\textsuperscript{th} Matthew Kelleher}
\IEEEauthorblockA{\textit{College of Medicine} \\
\textit{University of Cincinnati, OH }\\
Cincinnati, USA \\
kellehmw@ucmail.uc.edu}

\and
\IEEEauthorblockN{5\textsuperscript{th} Murat Ozer }
\IEEEauthorblockA{\textit{School of Information Technology} \\
\textit{University of Cincinnati, OH }\\
Cincinnati, USA \\
ozermm@ucmail.uc.edu}

\and
\IEEEauthorblockN{6\textsuperscript{th} Shane Halse}
\IEEEauthorblockA{\textit{School of Information Technology} \\
\textit{University of Cincinnati, OH }\\
Cincinnati, USA \\
halsese@ucmail.uc.edu}
}
\maketitle

\begin{abstract}

As Artificial Intelligence (AI) and Agentic AI become increasingly integrated across sectors such as education and healthcare, it is critical to ensure that Multi-Agent Education System (MAES) is explainable from the early stages of requirements engineering (RE) within the AI software development lifecycle. Explainability is essential to build trust, promote transparency, and enable effective human-AI collaboration. Although personas are well-established in human-computer interaction to represent users and capture their needs and behaviors, their role in RE for explainable MAES remains underexplored. This paper proposes a human-first, persona-driven, explainable MAES RE framework and demonstrates the framework through a MAES for clinical reasoning training. The framework integrates personas and user stories throughout the RE process to capture the needs, goals, and interactions of various stakeholders, including medical educators, medical students, AI patient agent, and clinical agents (physical exam agent, diagnostic agent, clinical intervention agent, supervisor agent, evaluation agent). The goals, underlying models, and knowledge base shape agent interactions and inform explainability requirements that guided the clinical reasoning training of medical students. A post-usage survey found that more than 78\% of medical students reported that MAES improved their clinical reasoning skills. These findings demonstrate that RE based on persona effectively connects technical requirements with non-technical medical students from a human-centered approach, ensuring that explainable MAES are trustworthy, interpretable, and aligned with authentic clinical scenarios from the early stages of the AI system engineering. The partial MAES for the clinical scenario simulator is~\href{https://github.com/2sigmaEdTech/MAS/}{open sourced here}.

\end{abstract}


\section{Introduction}

The integration of AI and Agentic AI into high-stakes domains such as healthcare and education is accelerating, with Multi-Agent Systems (MAS) offering sophisticated solutions for complex simulations and decision support~\cite{zhengFuzzySupervisorAgent2025}. In medical education, MAS can simulate realistic physician-patient encounters, enabling medical students to practice diagnostic and communication skills in a controlled learning environment~\cite{zhengUserCenteredIterativeDesign2025}. However, the increasing autonomy and complexity of these AI agents often result in ``black-box'' behavior~\cite{hassijaInterpretingBlackBoxModels2024}, where the reasoning behind their decisions is opaque to users. This lack of transparency poses a significant barrier to trust, user acceptance, and effective learning, as medical educators and students cannot verify or understand the AI's logic.

The central challenge is the absence of a systematic human-centered methodology for engineering requirements for eXplainable AI (XAI). Traditional Requirements Engineering (RE) processes are ill-equipped to handle the unique demands of data-driven AI systems, particularly the need to specify who requires an explanation, what needs explaining, and how the explanation should be delivered~\cite{habibaHowMatureRequirements2024}. Without a structured approach, the features of XAI are often developed ad-hoc, failing to meet the nuanced needs of diverse stakeholders. Habiba et al.~\cite{habibaCanRequirementsEngineering2022} highlight the importance of user-centered design in XAI and proposed a framework to integrate XAI requirements into RE processes by identifying stakeholders as a starting point, but there is a gap in methodologies that effectively translate user needs into technical requirements for explainability, especially in real-world use cases. Cirque et al.~\cite{cirqueiraScenarioBasedRequirementsElicitation2020} proposed a scenario-based requirement elicitation method for user-centric XAI with a fraud detection case study, but it does not explicitly address explainability in MAS, and the scenarios are not designed to capture the needs of multiple interacting agents in a complex system. There is a need for methodologies that can systematically capture and translate the explainability requirements of various stakeholders into actionable specifications for MAS in education.

This paper introduces a novel methodology that integrates Personas and XAI-specific User Stories into the RE of MAES development lifecycle. Personas are a concept from human-computer interaction design \cite{barambonesChatGPTLearningHCI2024}, which is not to represent a specific end-user, but as archetypal representations of a comprehensive group of "real" users and the AI agents themselves. This approach provides a RE framework for stakeholders to reason about the agents' internal states and decision-making processes. These personas are then used to generate XAI-focused user stories that explicitly capture explainability needs from the perspective of human users. 

Our primary contributions are threefold:
\begin{enumerate}
    \item A new methodology for XAI requirements engineering that utilizes AI Personas to model the explainability characteristics of AI agents.
    \item A practical AI Persona framework for authoring XAI-specific user stories that connect stakeholder goals to the internal behavior of a MAES.
    \item Validation of this methodology through a detailed case study of a MAES for Clinical Scenario Simulation (CSS), demonstrating its effectiveness in producing an explainable, trustworthy, and interpretable MAES.
\end{enumerate}

This paper is structured as follows: Section II introduces the background and related work in AI Agents and MAS, XAI in education, RE with personas. Section III details our proposed human-first, persona-driven, explainable MAES RE framework. Section IV presents a scenario simulator MAES for Clinical Reasoning Training. Section V discusses the results and their implications. Finally, Section VI concludes with key findings and directions for future work.

\section{Background and Related Work}

\subsection{AI Agents and Multi-Agent Systems}
Agents \cite{russell2021aima} are autonomous entities that perceive their environment and take actions to achieve specific goals. AI agents usually contain planning, tool calling, memory, and actions \cite{weng2023agent}. In MAS, multiple AI agents \cite{guoLargeLanguageModel2024} interact with each other and their environment, often collaborating or competing to solve complex problems.   The explainability of AI agents in MAS is particularly challenging due to their distributed nature and the emergent behaviors that arise from agent interactions \cite{madumal2019grounded}. Understanding how individual agents make decisions and how these decisions impact the system as a whole is crucial for building trust and ensuring accountability. In educational contexts, MAS can simulate realistic scenarios that provide rich learning experiences \cite{jiang2024ai, tsatsou2019adaptive,ganesan2025revolutionizing}, but the opacity of agent decision-making can hinder user engagement and learning outcomes. Therefore, integrating explainability into the design of AI agents in MAS is essential for their effective use in education.

\subsection{Explainable AI in Education}
Explainable AI (XAI) is a field of research and practice focused on addressing the lack of transparency in the black-box issue\cite{minh2022explainable} and making AI systems' decisions and predictions understandable to humans\cite{phillips2021four}. Explainability is critical in education, where AI systems are used to provide personalized learning experiences, assess student performance, and offer feedback \cite{chaushi2023explainable}. In medical education, for example, AI agents that simulate patient interactions or provide diagnostic support \cite{zhengUserCenteredIterativeDesign2025} must be able to explain their reasoning to medical students and educators. This transparency is essential for building trust in the system and ensuring that users can understand the AI's decision-making process. XAI techniques can help bridge the gap between complex AI models and human users by providing information on how decisions are made, which features are the most influential, and what uncertainties exist in predictions \cite{csahin2025unlocking}. This is particularly important in educational settings, where the goal is not only to achieve accurate results but also to facilitate learning and understanding \cite{saurel2025maestro}. Some research efforts have explored XAI in education \cite{fiok2022explainable, alibayeva2025data}, but there is still a need for systematic methodologies to integrate explainability into the design and development of educational AI systems, especially in multi-agent contexts where multiple AI entities interact with each other and with human users. Existing XAI techniques often focus on post-hoc explanations of model behavior\cite{zhang2025multi}. However, the application of XAI in educational contexts, particularly in multi-agent systems, is still an emerging area of research. There is a need for methodologies that can systematically capture and translate the explainability requirements of various stakeholders into actionable specifications for educational AI systems.

\subsection{Requirement Engineering and Personas}
Requirements Engineering (RE) is the process of defining, documenting, and maintaining  requirements for a computer-based system \cite{curcio2018requirements}. It is a critical process that ensures that systems meet user needs and expectations on what a system should do and how it should behave. The success of this RE stage could save costly rework and lay the foundation for project success \cite{sommerville1997viewpoints}. Nuseibeh and Easterbrook regard RE as a socio-technical activity that should involve various stakeholders and understand their perspectives on a RE roadmap \cite{nuseibeh2000requirements}. For AI systems, RE is particularly challenging due to their data-driven nature, probabilistic outputs, and emergent behaviors \cite{belani2019requirements, gjorgjevikj2023requirements}. Traditional RE methods often fall to capture these dynamic aspects. It is crucial to have a structured approach to elicit and specify requirements that address the unique challenges of AI systems, including explainability\cite{liywalii2024requirements}. Personas, fictional characters created to represent the different types of users within a targeted demographic, are a widely adopted technique in human-computer interaction for user-centered design to foster empathy and guide development\cite{zhengUserCenteredIterativeDesign2025, jansen2022create}. Currently, personas help teams focus on user goals and needs \cite{canedo2023use}. However, most applications have been limited to human users. The concept of applying personas to non-human entities, such as AI agents, to explore system interaction requirements remains largely unexplored. Creating detailed personas for each AI agent in a multi-agent system can help users better understand and communicate their capabilities, limitations, and decision-making processes. This approach allows stakeholders to reason about the agents' internal states and how they interact with human users, which is essential for eliciting explainability requirements that are aligned with user needs and expectations.

\section{Methodology: Persona-Based RE for Explainable MAES}
\begin{figure}[htbp]
\centerline{\includegraphics[width=0.85\columnwidth]{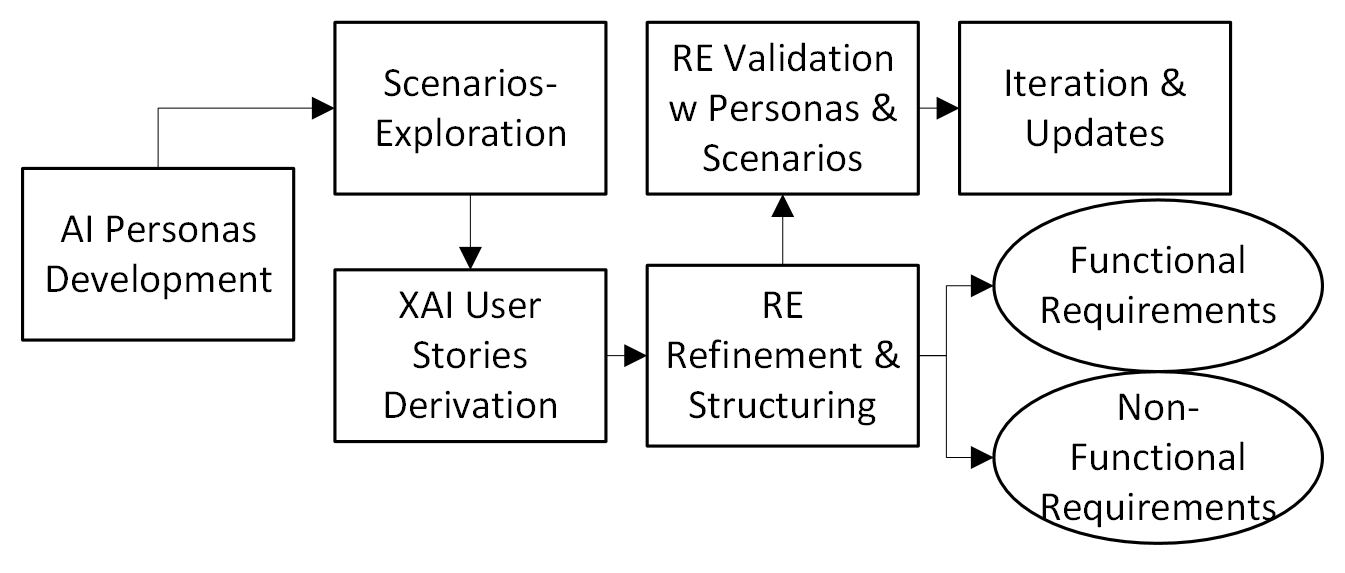}}
\caption{Persona-Based RE Framework for Explainable MAES}
\label{fig1:REFramework}
\end{figure}

Our methodology embeds the principles of user-centered persona-based design into the technical fabric of XAI requirements engineering. As shown in Figure~\ref{fig1:REFramework}, it consists of six core steps: AI persona development, scenario exploration, XAI User Stories Derivation,  RE refinement \& structuring, RE validation with Personas and Scenarios, RE Iteration and Updates.

\subsection{AI Personas Development}
We redefine the traditional persona concept to represent the AI agents within the MAES. An AI Persona is a semi-fictional archetypal representation of an AI agent, designed to make its capabilities, limitations, and decision-making processes tangible to stakeholders. AI agents could contain various components, such as the underlying model architecture or underlying tools and functions, the knowledge base, decision triggers, and the explainability profile. Creating detailed personas for each AI agent is the first step toward better understanding and communicating their internal workings and how they interact with human personas. This approach allows stakeholders to reason about the agents' internal states and decision-making processes, which is essential for eliciting explainability requirements that are aligned with user needs and expectations. The key attributes of an AI Persona include:
\begin{itemize}
    \item \textbf{Role \& Goal:} Its primary function within the MAES (e.g., ``AI patient'').
    \item \textbf{Model Architecture:} The underlying model, tools, or functions (e.g., ``Patient Simulation Engine'').
    \item \textbf{Knowledge Base:} The data on which it was trained and has access (e.g., ``MIMIC-IV dataset, patient vitals'').
    \item \textbf{Decision Triggers:} The inputs that prompt it to act.
    \item \textbf{Explainability Profile:} Its inherent capacity for explanation (e.g., ``Can provide feature importance but not counter factuals'').
\end{itemize}

\subsection{Scenarios Exploration}
After developing AI Personas, we explore various scenarios that represent typical interactions between human personas and AI agent personas. These scenarios are designed to capture the context in which explainability is the most critical. For example, in a clinical scenario simulation, a scenario might involve a medical student interacting with an AI patient agent to diagnose a condition. The scenario would detail the sequence of interactions, the information exchanged, and the decision points where explainability is essential for the student to understand the AI's reasoning. This exploration helps identify the specific moments where explainability is needed and informs the subsequent generation of XAI user stories.

\subsection{XAI User Stories Derivation and Prioritization}
With the AI Personas established and scenarios explored, we generate user stories that explicitly target explainability. We adapt the conventional user story template to focus on the need for understanding using the following structure.

\textit{``As a \textbf{[Human Persona]}, I want to understand \textbf{[Why/What/How]} the \textbf{[AI Persona]} made a \textbf{[Decision/Prediction/Recommendation]}, so that I can \textbf{[Achieve a Goal/Build Trust/Take Action]}."}

For example: \textit{``As a \textbf{Medical Student}, I want to understand \textbf{why} the \textbf{AI patient} ruled out a specific patient case, so that I can \textbf{learn its clinical symbolism for a special condition}."}

These user stories are then prioritized based on factors such as clinical risk, learning objectives, and implementation complexity.

\subsection{RE Refinement \& Structuring}

The prioritized XAI user stories become first-class citizens in the RE process. They are translated into formal functional and non-functional requirements. For instance, the user story above might translate to:
\begin{itemize}
    \item \textbf{Functional Requirement:} The system shall display the main features that contributed negatively to the ruled-out diagnosis.
    \item \textbf{Non-Functional Requirement:} The explanation shall be rendered within 500 ms and be understandable to a user with a second-year medical student's knowledge.
\end{itemize}
This ensures that explainability is not an afterthought,  but a core testable feature of the system.

\subsection{RE Validation with Personas and Scenarios}
RE validation is conducted through walkthroughs of the AI Personas and scenarios with human personas. This process ensures that the requirements derived from the user stories are aligned with the needs and expectations of the human personas. Stakeholders and human personas can provide feedback on whether the explainability features meet their needs and suggest adjustments to the requirements as necessary.

\subsection{Iteration \& Updates}
The RE process is iterative. As MAES is developed and tested, new insights may emerge that require updates to the AI Personas, scenarios, user stories, and requirements. This iterative approach ensures that the system remains aligned with user needs and can adapt to changes in the educational context or technological capabilities.

\section{Multi-Agent Educational System (MAES) for Clinical Scenario Simulator (CSS): A Case Study}
We applied our methodology to the development of a MAES for CSS in medical education for clinical reasoning training.

\subsection{Identifying High Level Architecture of MAES for CSS}
\begin{figure}[htbp]
\centerline{\includegraphics[width=0.75\columnwidth]{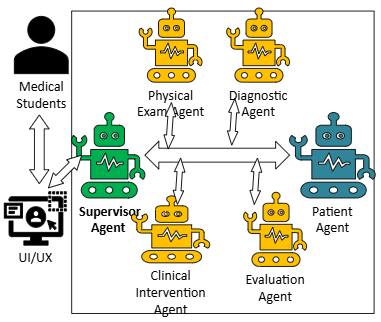}}
\caption{AI Agents and their Relationships in MAES for CSS}
\label{fig2:Agents}
\end{figure}

The MAES for CSS platform is designed to simulate authentic doctor-patient encounters for medical students. Medical students act as doctors. The MAES builds a CSS and acts as patients and clinical agents for clinical reasoning training environments. The simulator provides different clinical cases for different levels of medical students to practice their clinical reasoning skills~\cite{zhengUserCenteredIterativeDesign2025}. The relationships between different agents are shown in Figure~\ref{fig2:Agents}. The MAES is orchestrated by the supervisor agent with several autonomous clinical agents to setup the clinical scenario simulator\cite{zhengFuzzySupervisorAgent2025}. From top to bottom, medical students log into the platform UI and choose a simulated clinical case. The message of starting a new case passes through the supervisor agent to initialize all other agents.  Each agent represents a distinct aspect of the MAES for CSS workflow: the Patient Agent simulates patient responses and symptoms; the Physical Exam Agent manages examination findings; the Diagnostic Agent handles test ordering and results; the Clinical Intervention Agent oversees therapeutic actions; the Evaluation Agent assesses overall student performance and generates the final feedback report; the supervisor agent manages all other agents and provides real-time responses. Medical students interact with these agents through a unified user interface (UI) and generate rich interaction and behavioral data, such as questions asked, ordered tests, and performed interventions.

\subsection{Identifying Stakeholders, AI Agents and their Roles in MAES for CSS} 

MAES for CSS has multiple stakeholders, including medical students, medical educators, and the AI simulated agents. Each stakeholder has unique needs and goals that have been considered in the design of the system. Medical students require a realistic and interactive learning environment to practice clinical reasoning skills. Medical educators need tools to monitor student progress and provide targeted feedback. AI agents are designed to simulate realistic patient interactions and provide accurate diagnostic support while being explainable in their decision-making processes. The following table \ref{tab:Stakeholders} summarizes the key stakeholders and their roles in the MAES for CSS. This comprehensive understanding of stakeholders and their interactions is crucial for developing explainability requirements that align with the needs of all parties involved in the clinical scenario simulator. 

{\small
\begin{table}[htbp]
\caption{Stakeholders and Roles in MAES for CSS}
\label{tab:Stakeholders}
\begin{center}
\begin{tabular}{|p{3.0cm}|p{3.8cm}|}
\hline
\textbf{Stakeholder} & \textbf{Role} \\
\hline
Medical Students & Primary users to train clinical reasoning skills. \\
\hline
Medical Educators & Monitor student progress and provide feedback. \\
\hline
AI Patient Agent & Simulates patient responses and symptoms for clinical case. \\
\hline
AI Physical Exam Agent & Manages examination findings and provides feedback on physical exam performance. \\
\hline
AI Diagnostic Agent & Handles test ordering, results, and provides diagnostic support. \\
\hline
AI Clinical Intervention Agent & Oversees therapeutic actions and provides feedback on intervention choices. \\
\hline
AI Evaluation Agent & Assesses overall student performance and generates feedback. \\
\hline
AI Supervisor Agent & Orchestrates the simulation and provides real-time responses. \\
\hline
\end{tabular}
\end{center}
\end{table}
}
\subsection{Creating AI Personas in MAES for CSS}

Following key attributes, AI Personas were designed to capture unique goals, models, knowledge bases, decision triggers, and explainability profiles. The Decision Triggers are triggered by medical students by entering the needs. The following are the AI Personas for each agent in the MAES for CSS:

{\small
\textbf{Name:} Alex (AI Patient Agent)
\begin{itemize}
    \item \textbf{Goal:} Provide realistic patient interactions and symptoms based on the clinical case without revealing the underlying diagnosis prematurely to medical students.
    \item \textbf{Model:} Large language model-based dialogue agent guided for clinical conversations and a Patient Simulation Engine for symptom generation.
    \item \textbf{Knowledge Base:} Clinical case scripts and structured templates.
    \item \textbf{Explainability:} Can provide interaction-level explanations based on the history of questions asked.
\end{itemize}

\textbf{Name:} Dr.\ Eva (AI Physical Exam Agent)
\begin{itemize}
    \item \textbf{Goal:} To manage the findings of the physical examinations and provide feedback on the performance of the physical examinations.
    \item \textbf{Model:} Rule-based exam reasoning engine and an interface for student interaction to accept exam requests and return structured findings.
    \item \textbf{Knowledge Base:} Tables for findings of clinical case specific exam (e.g. vital signs).
    \item \textbf{Explainability:} Can provide procedural explanations that highlight the specific areas of the coverage and findings of physical exams. 
\end{itemize}

\textbf{Name:} Brian (AI Diagnostic Agent)
\begin{itemize}
    \item \textbf{Goal:} Handle investigation ordering and result reporting, and support diagnostic reasoning interpretation.
    \item \textbf{Model:} Rule-based test catalog and diagnostic reasoning engine.
    \item \textbf{Knowledge Base:} A structured test catalog: laboratories, imaging, and procedures, and encoded disease-test evidence.
    \item \textbf{Explainability:} Can provide test-utility explanations and show the probabilistic links between post-test results and diseases.
\end{itemize}

\textbf{Name:} Clair (AI Clinical Intervention Agent)
\begin{itemize}
    \item \textbf{Goal:} To oversee therapeutic actions and provide feedback on intervention choices (medications, procedures, supportive care) and simulate the results.
    \item \textbf{Model:} Rule-based intervention treatment protocol engine.
    \item \textbf{Knowledge Base:} Student interaction logs, treatment protocol guidelines, and clinical pathways for common conditions.
    \item \textbf{Explainability:} Can provide guideline-based rationales and flag safety concerns with reason codes.
\end{itemize}

\textbf{Name:} Dr.\ Eval (AI Evaluation Agent)
\begin{itemize}
    \item \textbf{Goal:} To aggregate and assess overall student-agent conversation logs and generate structured feedback reports.
    \item \textbf{Model:} Scoring and analytic engine combined with a rule-based scoring rubrics and reasoning large language model to analyze student's performance and provide feedbacks.
    \item \textbf{Knowledge Base:} Student interaction logs, medical educator's defined assessment rubrics, and competency frameworks. 
    \item \textbf{Explainability:} Can provide explanations for its rubric-based evaluation decisions and highlight the key factors that influenced its evaluation.
\end{itemize}

\textbf{Name:} Sam (AI Supervisor Agent)
\begin{itemize}
    \item \textbf{Goal:} To orchestrate the entire simulation, manage state, and coordinate all AI agents and student interactions.
    \item \textbf{Model:} An orchestration engine that manages the flow of information and interactions between agents and students, with a routing logic based on rule-based decision systems for real-time adjustments.
    \item \textbf{Knowledge Base:} Student interaction logs, agent status, and predefined orchestration rules.
    \item \textbf{Explainability:} Can explain scenario-flow decisions and display case progression logic.
\end{itemize}
}

\subsection{Exploring Clinical Scenarios}

We explore various clinical scenarios that medical students might encounter while using the MAES for CSS. For example, in a clinical case, a medical student could involve an AI patient (Alex) who presents with chest pain, where the AI Diagnostic Agent (Brain) must analyze symptoms and test results to generate a differential diagnosis. The AI Supervisor Agent (Sam) would then provide feedback based on the student's conversation about collecting the patient symptoms and clinical conditions. These scenarios were designed to be realistic and educational, allowing students to practice their clinical reasoning skills while also providing opportunities for AI agents to demonstrate their explainability features. For example, in the chest pain scenario, Sam might explain that it ruled out a diagnosis of myocardial infarction because the patient's EKG results did not show specific changes, and Dr. Eva might provide feedback on the student's choice of diagnostic tests, explaining that it recommended a troponin test due to the patient's risk factors and presentation. Finally, Dr.Eval will collect all the conversation between student-AI agents and provide structured feedback on what could be improved and what did good. Clinical scenarios could help identify the interaction and relationship between human personas and AI personas, and we could explore as much as we needed. These clinical scenarios help identify the specific moments where explainability is needed and inform the subsequent generation of XAI user stories. 

\subsection{Identifying XAI User Stories for each stakeholder and agent}
Based on the identified AI agents and clinical scenarios, we develop user stories that capture the explainability requirements for each stakeholder and agent in the MAES for CSS. These user stories were generated according to a combination of stakeholder interviews, focus groups, and LLM-assisted brainstorming sessions \cite{zhengUserCenteredIterativeDesign2025}. The user stories are designed to ensure that the system's explainability features are aligned with the needs and expectations of all stakeholders and AI agents. For example, a user story for the AI Diagnostic Agent might be: \textit{``As a medical student, I want to understand why the AI Diagnostic Agent ruled out a specific diagnosis, so that I can learn from its reasoning process and improve my clinical decision-making skills.''} This user story highlights the need for transparency in the diagnostic process and informs the requirement for the AI Diagnostic Agent to provide clear explanations of its reasoning. Another user story for the AI Supervisor Agent could be: \textit{``As a medical educator, I want to understand the criteria the AI Evaluation Agent uses to provide feedback, so that I can better guide my students and ensure that the feedback is educationally valuable.''} This user story emphasizes the importance of explainability in the feedback mechanism, ensuring that educators can trust and effectively utilize the system's feedback to enhance student learning outcomes. These user stories serve as a foundation for translating stakeholder needs into specific actionable requirements that guide the development of explainability features in the MAES for CSS.

\subsection{Translating User Stories into Requirements}
The user stories were then translated into specific functional and non-functional requirements for the MAES for CSS. For example, the user story about understanding the AI Diagnostic Agent's reasoning process might translate into a functional requirement that states: \textit{``The AI Diagnostic Agent shall provide a feature importance visualization that highlights the symptoms and test results that most strongly influenced its top diagnosis.''} This requirement ensures that the system includes a specific explainability feature that allows medical students to understand the diagnostic reasoning of the AI Diagnostic Agent. Similarly, the user story about understanding the AI Evaluation Agent's feedback criteria might translate into a requirement such as: \textit{``The AI Evaluation Agent shall log its feedback decisions and the triggers for those decisions in an educator dashboard, allowing educators to review and understand its pedagogical strategy.''} This requirement ensures that educators have access to information about how the evaluation agent is making its feedback decisions, which is essential for building trust and effectively using the system in an educational context. These requirements were then integrated into the development process of the MAES for CSS, ensuring that explainability was a core aspect of the system's design and functionality from the early stages of development.

\subsection{Validating RE with AI Personas and Scenarios}
After translating the user stories into requirements, we validated these requirements through walkthroughs of the AI Personas and scenarios with human agents. This process involved presenting AI Personas and clinical scenarios to medical students and educators, along with the derived requirements for explainability features. Stakeholders were asked to provide feedback on whether  requirements met their needs and expectations for explainability in the context of the clinical scenario simulation. For example, medical students might have provided feedback on whether the feature importance visualization (a clinical reasoning decision tree) for the AI Diagnostic Agent was clear and helpful in understanding diagnostic reasoning, while educators might have evaluated whether the educator dashboard for the AI Evaluation Agent provided sufficient insight into its feedback decisions. This validation process ensured that the requirements were aligned with the needs of the stakeholder and informed any necessary adjustments to better meet those needs.

\subsection{Iterating and Updating RE}
The RE process was iterative, with ongoing feedback from stakeholders that led to updates in AI Personas, scenarios, user stories, and requirements. As MAES for CSS was developed and tested, new insights emerged that prompted revisions to the explainability features. For example, if medical students found that the initial feature importance visualization was too complex or not sufficiently informative, we might have iterated on the design of that feature to make it more user-friendly and educationally valuable. Similarly, if educators provided feedback that the educator dashboard did not capture certain aspects of the supervisor agent's decision-making process, we could have updated the requirements to include additional logging or visualization features. This iterative approach ensured that the MAES for CSS remained aligned with user needs and could adapt to changes in the educational context or technological capabilities as development progressed.

\section{Results and Discussion}
\subsection{Impact on Requirement Quality}
The use of AI Personas and XAI user stories led to requirements that were more specific, comprehensive, and user-focused. The persona ``Alex'' allowed stakeholders to articulate their need for transparency in a way that ``the diagnostic model'' did not. This AI persona RE framework created an effective bridge between human needs and the technical reality of MEAS. The resulting requirements covered a wider range of explainability needs than our initial expert-driven brainstorming had produced.

\subsection{Stakeholders Feedback}
We conducted a post-use survey with 42 medical students and 2 medical educators.
\begin{itemize}
    \item \textbf{Educational Value:} When asked if the app aligned with the clinical skills course objectives, 78.57\% of the students responded “agree” or “strongly agree”. 9.52\% of the 42 responses are “neutral”. That means that the majority of the students found the app to be a valuable educational tool that supports their clinical reasoning learning course objectives.
    \item \textbf{Clinical Reasoning Development:} When asked which aspects of the clinical scenarios' simulator were most valuable for clinical reasoning learning, 36 answers returned. 26 responses contain Clinical reasoning development, followed by 17 responses with real-time feedback, and 16 responses with history taking practice, while physical exam got 8 responses and independent learning pace has 7 responses.
    \item \textbf{Trust and Transparency:} When asked about the features that are most useful in the simulator, the survey collected 26 responses. All of them show that the simulator is helpful in their study. 10 responses respond to lab ordering with immediate results and the intervention tab is helpful, while 9 responses mentioned that it is helpful for their extra practice for clinical reasoning learning, and 5 answered that “Talk to supervisor” is very helpful. The feedback feature received 5 votes. These results suggest that the explainability of each agent features, such as immediate test results and feedback, is highly valued by students for their learning experience.
\end{itemize}

\subsection{Challenges and Limitations}
The major challenge in this approach was to ensure that the AI Personas were grounded in the technical capabilities of the underlying models. While the personas were designed to be relatable and human-like, it was essential to maintain a clear distinction between the persona's attributes and the actual technical specifications of the AI agents. This required careful facilitation during stakeholder discussions to prevent over-anthropomorphizing the AI agents, which could lead to unrealistic expectations about their capabilities. Additionally, the quality of the user stories and the requirements derived from the personas were highly dependent on the quality of the input data and the prompts used to generate the personas. If the initial data or prompts were biased or incomplete, it could lead to personas that do not accurately represent the AI agents, which, in turn, could result in requirements that are not aligned with user needs. Also, because of limited case studies, more detailed scenario and XAI user stories, as well as RE, could be improved and enhanced. Finally, this case study is limited to a single application in medical education, and further research is needed to validate this methodology in other domains and with different types of AI systems.

\section{Conclusion and Future Work}
This paper has demonstrated that the integration of AI Personas and XAI-specific user stories provides a robust methodology for requirements engineering in explainable AI systems. Treating AI agents as first-class stakeholders in the design process, we can systematically elicit and specify requirements for explainability, transparency, and interpretability. Our case study of a multi-agent clinical simulator confirms that this human-centered, persona-based RE for MEAS approach results in systems that are not only more trustworthy but also more effective in achieving their educational goals.

For future work, we plan to explore more detailed scenarios and XAI user stories, as well as functionality and non-functional requirements. The underlying models, tools, and functional calls in each agent could be improved. We will also investigate the application of this methodology in other high-stakes domains, such as K-12 educational settings, literature reviews, to develop a generalized framework for human-centered XAI design. Finally, we will examine the ethical implications of using anthropomorphic personas in AI development, particularly with respect to user expectations and the potential for deception.

\section*{Acknowledgment}
{\small
The manuscript and GitHub code used BearcatGPT to polish and help. The authors thank all medical students and experts for their participation and invaluable feedback.}

\bibliographystyle{IEEEtran}
\bibliography{references.bib}

\end{document}